\documentclass[conference]{IEEEtran}
\IEEEoverridecommandlockouts
\usepackage{cite}
\usepackage{amsmath,amssymb,amsfonts}
\usepackage{algorithmic}
\usepackage{graphicx}
\usepackage{epstopdf}
\usepackage{textcomp}
\usepackage{xcolor}
\usepackage{url} 
\usepackage{listings}
\usepackage{marvosym}
\usepackage{enumitem}
\usepackage{pifont}
\usepackage{multirow}
\usepackage{cleveref}
\usepackage{subfig}
\def\BibTeX{{\rm B\kern-.05em{\sc i\kern-.025em b}\kern-.08em
    T\kern-.1667em\lower.7ex\hbox{E}\kern-.125emX}}
    
\begin{document}

\title{ RARO: Reliability-Aware Read Optimization for Hybrid Flash Storage}

\author{
\IEEEauthorblockN{
    Yanyun Wang\textsuperscript{1}\thanks{Equal contribution}, Dingcui Yu\textsuperscript{1}\thanks{Equal contribution},
    Yina Lv\textsuperscript{2}, Yunpeng Song\textsuperscript{1},
    Yumiao zhao\textsuperscript{1}, 
    Liang Shi\textsuperscript{1}
} 

\IEEEauthorblockA{\textsuperscript{1}College of Computer Science and Technology, East China Normal University, Shanghai, China} 

\IEEEauthorblockA{\textsuperscript{2}School of Informatics, Xiamen University, Xiamen, China}

51265901050@stu.ecnu.edu.cn, \{dingcuiy, elainelv95\}@gmail.com, 
51205901045@stu.ecnu.edu.cn \\
zhaoyumiao99@gmail.com, shi.liang.hk@gmail.com 
}


\maketitle

\begin{abstract}
Quad-level cell (QLC) flash offers significant benefits in cost and capacity, but its limited reliability leads to frequent read retries, which severely degrade read performance.
A common strategy in high-density flash storage is to program selected blocks in a low-density mode (e.g., SLC), sacrificing some capacity to achieve higher I/O performance.
This hybrid storage architecture has been widely adopted in consumer-grade storage systems.
However, existing hybrid storage schemes typically focus on write performance and rely solely on data temperature for migration decisions.
This often results in excessive mode switching, causing substantial capacity overhead.

In this paper, we present RARO (\underline{R}eliability-\underline{A}ware \underline{R}ead performance \underline{O}ptimization), a hybrid flash management scheme designed to improve read performance with minimal capacity cost.
The key insight behind RARO is that much of the read slowdown in QLC flash is caused by read retries.
RARO triggers data migration only when hot data resides in QLC blocks experiencing a high number of read retries, significantly reducing unnecessary conversions and capacity loss.
Moreover, RARO supports fine-grained multi-mode conversions (SLC–TLC–QLC) to further minimize capacity overhead.
By leveraging real-time read retry statistics and flash characteristics, RARO mitigates over-conversion and optimizes I/O performance.
Experiments on the FEMU platform demonstrate that RARO significantly improves read performance across diverse workloads, with negligible impact on usable capacity.

\end{abstract}


\footnotetext[1]{First Author and Second Author contribute equally to this work.}

\section{Introduction}
As 3D NAND vertical stacking technology and multi-bit technology develop, flash memory technology has realized a leapfrog evolution from Single-Level Cell (SLC) to Penta-Level Cell (PLC). 
Among them, Quad-Level Cell (QLC) flash memory, by storing four bits per cell, which significantly increases the storage density and reduces the cost, has been widely used in consumer-grade storage systems. 

However, the increased bit density in QLC leads to narrower noise margins and more severe program interference and read disturb as layer counts increase.
As a result, QLC suffers from reduced endurance and notably higher read latency due to frequent read retries—a process that adjusts read reference voltages to correct errors when ECC (e.g., LDPC) fails.
A common strategy in consumer-grade flash storage is to combine the high-density flash with high-performance flash\cite{luo-hyghdensity,ssd665p,crucialP1} and achieve the trade-off between performance and cost.
Specifically, hybrid flash architectures are achieved by programming flash blocks into low-density modes (e.g., SLC or TLC) to improve I/O performance at the cost of shrinking capacity.
For instance, the Intel 665P \cite{ssd665p} employs a block-level pSLC strategy to reconfigure QLC blocks into SLC mode, which in turn improves read performance. 

Current management strategies for hybrid flash storage show multi-dimensional innovations. 
HyFlex \cite{luo2023performance} proposes a capacity tuning method based on data write speed and garbage collection sensing, which better optimizes data placement and flash mode management strategies. 
HAML-SSD \cite{8942140} adopts a two-dimensional clustering algorithm, which ensures that flash modes are aligned with dynamic workload variations by combining thermal data information, and significantly reduces the risk of performance degradation.
RLcSSD \cite{wei2023reinforcement} further improves the performance of hybrid SSDs by dynamically optimizing flash mode transitions and garbage collection through reinforcement learning. 
Hook \cite{10693661} utilizes a lightweight temporal and spatial location-aware prediction method to enable prefetching of SLC buffers to improve read performance without sacrificing write performance.

Despite these advances, three key limitations remain: 
1) \textbf{Lack of read-centric optimization}. 
Most of the existing research focuses on improving the write performance of flash memory through the write acceleration feature of low-density flash memory, while the research on the improvement of read performance is relatively weak.
2) \textbf{Excessive mode switching}.
Common mode switching strategies rely solely on data temperature for mode conversion and data migration, which often leads to excessive mode switching and significant flash capacity loss.
3) \textbf{Limited mode granularity}. 
Most of the existing mainstream schemes are SLC-QLC or SLC-TLC dual-mode switching. 
This coarse granularity not only causes significant capacity reduction but also leads to large fluctuations in I/O performance.

To address these issues, we propose RARO, a Reliability-Aware Read performance Optimization scheme for QLC SSDs. 
The key insight behind RARO is that much of the read slowdown in QLC flash is caused by read retries.
Specifically, RARO integrates four key modules into the SSD’s Flash Translation Layer (FTL): a heat classifier, RBER calculator, read retry counter, and flash mode translation controller.
RARO dynamically manages data migration and flash mode switching (SLC-TLC-QLC) based on both data hotness and cell reliability. 
By setting tiered read retry thresholds, it intelligently migrates hot data to more reliable, lower-density modes to improve read performance while minimizing capacity loss from excessive migrations. 
Additionally, an elastic capacity recovery policy is incorporated to maintain stable operation under varying workloads.
This significantly improves read performance while minimizing capacity loss.

RARO is implemented in FEMU\cite{li2018case}, a QEMU-based NVMe SSD emulator.
All blocks are initially set as QLC, and SLC/TLC blocks appear only after mode conversions.
To evaluate the effectiveness of RARO, we compared it against two strategies: the Baseline, which applies multiple read retries on QLC without mode awareness, and Hotness, a temperature-based mode conversion and migration scheme.
Workloads were generated with FIO \cite{Fio} using various Zipf-distributed random reads.
We evaluated the performance improvements of RARO under different wear levels.
Compared to the Baseline, random read IOPS increased by 9.3 to 14.25 times.
Compared to Hotness, RARO achieved similar IOPS performance while significantly reducing capacity loss.

The major contributions of this work are as follows:
\begin{itemize}
\item We analyze read retry behavior across flash modes and identify its impact on read latency, motivating a retry-aware mode management approach.
\item We propose a migration policy based on both flash reliability and data hotness to selectively switch storage modes, reducing unnecessary mode conversions and preserving storage capacity.
\item We simulate the proposed RARO approach using workloads with various access skewness.
Experimental results show that RARO significantly improves read performance with minimal capacity overhead.
\end{itemize}

\section{Backgroud}

\subsection{\textbf{3D NAND Flash Memory}}
Compared to planar NAND architectures, 3D NAND flash provides greater capacity and higher storage density with vertical stacking and multi-bit technology. 
There are three basic operations in 3D NAND flash memory: read, program/write, and erase, where read and program operations are usually based on the page, and erase operations are usually based on the block.
Write/programming of a flash memory cell is achieved by applying a high programming voltage (e.g., greater than 20 V) to the wordlines and bitlines of the target flash memory cell. 
The erase operation decreases the $V_{TH}$ level of the target cell by exerting a high erase voltage $V_{ERASE}$ on the substrate.

In 3D QLC flash memory, data read determines whether the data stored in a flash cell is a 1 or a 0 by comparing the voltage of the target flash cell with the read reference voltage.
For example, reading all four bits in QLC requires 15 read voltages $R_1 \sim R_{15}$ to determine, where reading a specific bit requires $1 \sim 8$ reads, depending on the QLC Gray code.

\subsection{\textbf{Hybrid SSD}}
Depending on the number of bits stored in one memory cell, they can be categorized as SLC (single-level-cell), MLC (multiple-level-cell), TLC (triple-level-cell), QLC (quad-level-cell), PLC (penta-level-cell) and so on.
For example, there are 4 bits in a QLC flash cell through 16 threshold voltage (Vth) levels, namely LSB (least-significant-bit), CSB (center-significant-bit), MSB (most-significant-bit), and TSB (top-significant-bit) from low to high.

To balance the performance, cell density, storage capacity, and reliability advantages of the various types of flash memories above, hybrid SSDs have further emerged.
Different types of NAND flash memory (e.g., SLC, TLC, and QLC) have common primary structures, with the distinction being the way in which the threshold voltage is programmed and sensed. 
Therefore, by varying the threshold voltage and ISPP, these flash cells can be switched, which has been demonstrated, so that hybrid SSDs can achieve a trade-off between their strengths.

\begin{figure}[t]
    \centering
    \includegraphics[width=8cm]{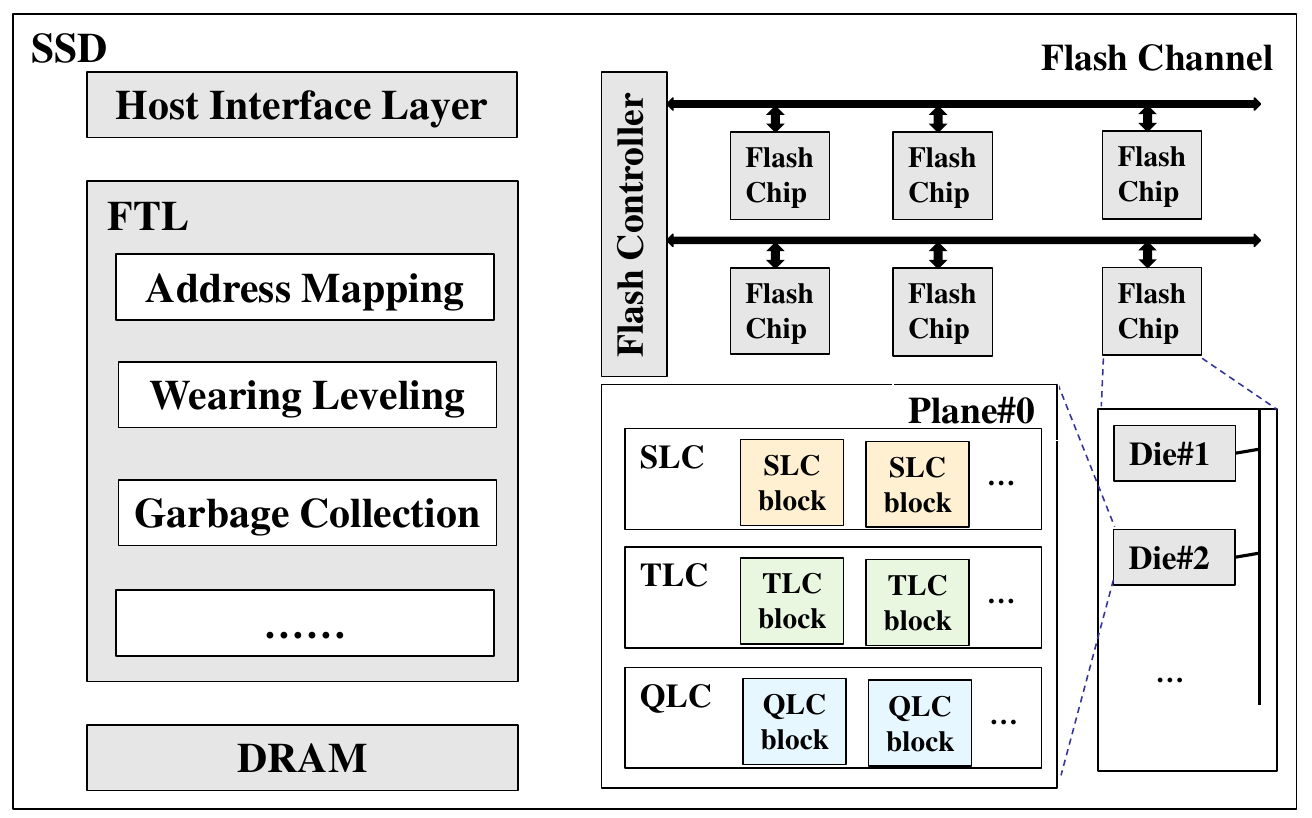}
    \caption{The architecture of QLC-TLC-SLC hybrid SSD.}
    \label{fig}
\end{figure}

\subsection{\textbf{Reliability Factors}}
3D NAND high-density flash memory is more susceptible to programming, erasure, and temperature impacts than 2D NAND, while the intensive arrangement of vertically oriented flash cells exacerbates cell-to-cell interference, and the unevenness of the etching process results in more pronounced layer-to-layer variations, leading to more severe reliability issues for 3D NAND flash memory \cite{7056062,10.1145/3224432,6657034,8327033}.
In this paper, we concentrate on three core reliability degradation factors - wear, retention loss, and read disturbances, then construct a multidimensional parametric model of the Raw Bit Error Rate (RBER) of flash memory:
\begin{equation}
\begin{aligned}[b]
\mathrm{\text { RBER }} 
& (\text { cycles }, \text { time }, \text { reads}) \\
& = \varepsilon +\alpha \cdot \text { cycles }^{k} &&\text{(wear)}\\
& +\beta \cdot \text { cycles }^{m} \cdot \text { time }^{n} && \text{(retention)} \\
& +\gamma \cdot \text { cycles }^{p} \cdot \text { reads }^{q} &&\text{(disturbance)} 
\label{eq}
\end{aligned}
\end{equation}

In Equation (1), wear refers to the fact that repeated programming and erasing operations (P/E cycle) can cause the tunnel oxide layer of the flash cell to thin. 
Data retention error refers to the gradual leakage of electrons stored in a flash memory cell over time, which may result in bit flipping and make it difficult to read data correctly. 
Read disturbance occurs when reading a wordline weakly programs other wordlines in the same block, injecting more electrons into its flash cell.
Consequently, these inherent characteristics make high-density flash memory particularly susceptible to various types of errors.

\subsection{\textbf{Retry Modeling based on RBER}}
Based on a previous work on reliability modeling \cite{kim2019design,lv2023mgc}, we can construct a read-retry model similar to Equation (2) based on the raw bit error rate (RBER) of a data page to ascertain the average number of read retries for each page in a specific phase.
\begin{equation}
\begin{aligned}[b]
\left(\alpha * RBER * n_{SENSE} \right) *(1-\delta)^ {n_{RETRY}} \leq E_{LDPC}
\label{eq}
\end{aligned}
\end{equation}
\begin{equation}
\begin{aligned}[b]
n_{RETRY} \geq \log _{(1-\delta)}\left(\frac{E_{LDPC}}{\alpha * RBER * n_{SENSE}}\right)
\label{eq}
\end{aligned}
\end{equation}

In Equation \eqref{eq}, the variable $\alpha$ reflects the RBER for two adjacent voltage states, and $ n_{SENSE} $ indicates the number of applied reference voltages.
For QLC flash memory, the number of read reference voltages required varies with the used Gray codes. 
Generally speaking, the higher the read levels, the more reference voltages are needed, and the worse the read performance.
$\delta$ indicates the raw error bit rate that can be reduced after each read retry. $(1-\delta)^ {n_{RETRY}}$ is the raw error bit rate that can be decreased after n retries. 
For example, a $\delta$ of 20\% means that the RBER of the read page can be dropped to 80\% of the original RBER after each retry.
$E_{LDPC}$ in this paper takes 72 to mean that 72 bits error can be corrected every 1KiB code-word.

\section{Motivation}
In this section, we conducted preliminary experiments to study the read performance of different flash memories and collect the number of read retries of TLC and QLC flash memories.
The evaluation environment and device parameters are described in Section V.
In preliminary experiments, the Flexible I/O Tester (FIO) \cite{Fio} is employed to generate some workloads, including Zipf. 

\subsection{\textbf{Comparison of Various Flash Modes}}\label{AA}
The total read data of the experimental workload is 8 GB, and the read performance of different modes of flash memory is tested under random read and sequential read, respectively.

\begin{figure}[htbp]
    \centering
    \includegraphics[width=\linewidth]{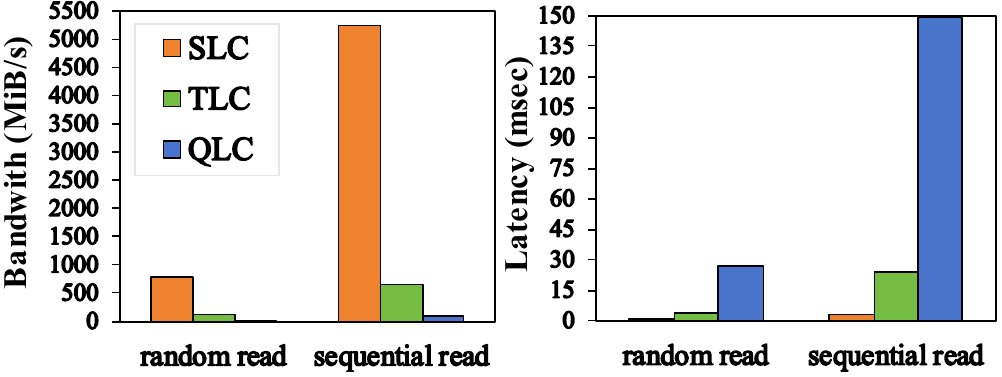}
    \caption{Read performance comparison among QLC, TLC, and SLC flash memories.}
    \label{fig2}
\end{figure}

As shown in \Cref{fig2}, the read performance of SLC flash, TLC flash, and QLC flash varies greatly in both random read 4K and sequential read 128K, with QLC degrading by about 63.6\% compared to SLC in the sequential read 128K scenario.

\subsection{\textbf{Read Retry induced Performance Impact}}
\begin{figure}[htbp]
    \centering
    \includegraphics[width=\linewidth]{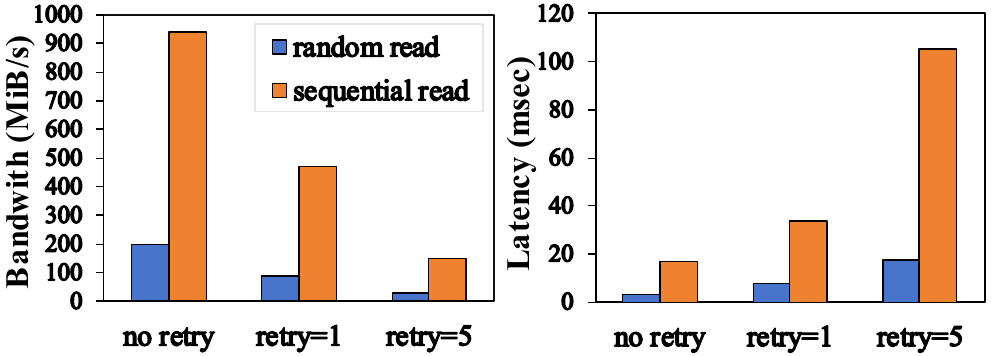}
    \caption{Read performance impact by varying the number of retry counts on TLC flash memory.}
    \label{fig3}
\end{figure}

\begin{figure}[htbp]
    \centering
    \includegraphics[width=\linewidth]{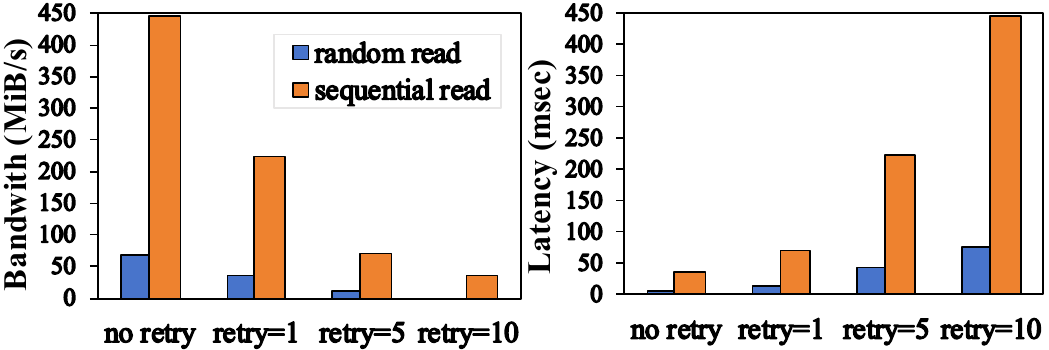}
    \caption{Read performance impact by varying the number of retry counts on QLC flash memory.}
    \label{fig4}
\end{figure}

To reflect the performance impact of the read retries, we first evaluate the read performance under fixed P/E cycles. 
\Cref{fig3} and \Cref{fig4} show the effect of varying the number of read retries on the performance of random and sequential accesses to $16KB$ of TLC and QLC flash memory, with the same temperature and P/E cycle. 
The non-retry is the first page read of 16KB data.
When the first page read fails to correct the data with ECC, a read retry operation is performed.

As shown in Figure \ref{fig4}, for QLC flash, one retry and ten retries reduce the sequential read bandwidth by 50.0\% and 92.0\%, respectively, and the random read bandwidth by 46.5\% and 90.5\%, respectively, compared to no read retries.

\subsection{\textbf{Distribution of Read Retry}}
Prior work \cite{cho2024aero,park2021reducing,lv2023mgc} has found that flash reliability and P/E cycles are linearly related, so we divide TLC and QLC into three reliability stages based on the P/E cycles as shown in Table \ref{tab:reliability}.


\begin{table}[htbp]
\caption{Three Reliability Stages for QLC}
\begin{center}
\begin{tabular}{|c|c|c|c|}
\hline
\textbf{Reliability stage} & \textbf{Young} & \textbf{Middle} & \textbf{Old} \\
\hline
\textbf{\textit{P/E cycle}} & $0\sim333$ & $334\sim666$ & $667\sim1000$ \\
\hline
\end{tabular}
\end{center}
\label{tab:reliability}
\end{table}

\begin{figure}[htbp]
    \centering

    \includegraphics[width=7cm]{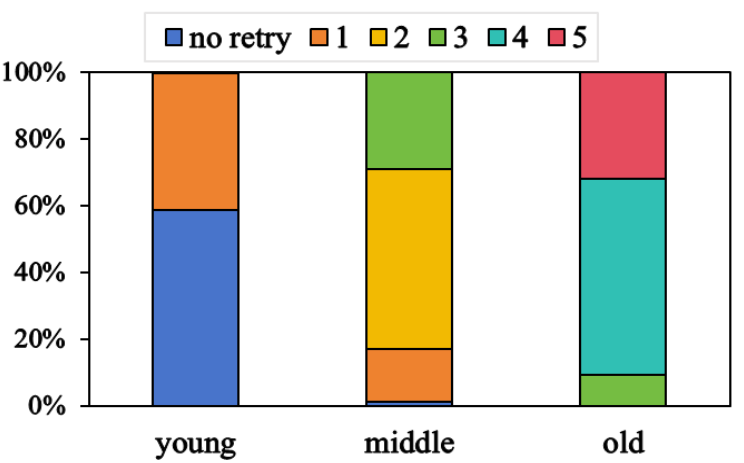}
    
    
    \caption{The number of page retries distribution at young/middle/old stages.}
    \label{fig5}
\end{figure}

\begin{figure}[htbp]
    \centering
    \includegraphics[width=8.5cm]{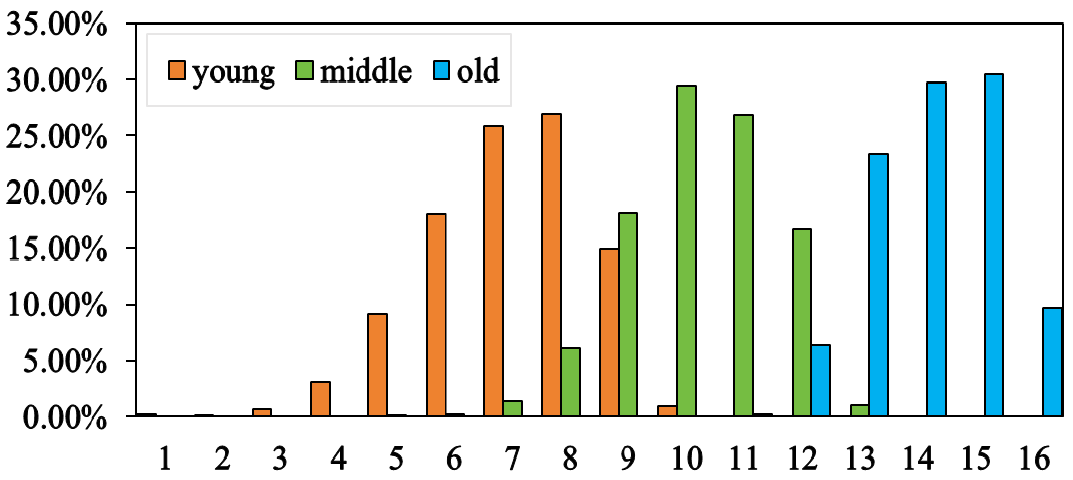}
    \caption{The number of page retries distribution in QLC young/middle/old stages. Young stage refers to the P/E cycle between 0 and 333, middle age refers to the P/E cycle between 334 and 666, and old age refers to the P/E cycle between 667 and 1000.}
    \label{fig6}
\end{figure}
\Cref{fig5} shows the page read retry distribution of TLC flash and
\Cref{fig6} shows the variation of read retries of QLC flash memory from the young stage to the old stage, respectively.
As shown in \Cref{fig6}, the read retries in the young stage are roughly $1\sim10$, in the middle stage are roughly $5\sim13$ retries, and in the old stage are roughly $11\sim16$ read retries.
Several findings can be concluded: 
1) A large number of read retries may occur in modern 3D NAND flash memory, and the more retries, the greater the performance degradation. 
2) As the reliability of the flash memory decreases, the number of retries required gradually increases. 
3) The retries of QLC are much more severe compared to TLC at the same stage; 
4) The percentage of pages in which the maximum number of read retries occurs is relatively small in different reliability stages of flash memory. 
Take QLC for example, the maximum number of read retries in the young stage is 10, accounting for 1.01\% of the total number of pages; the maximum number of retries in the middle stage is 13, accounting for 1.04\%; and the maximum number of retries in the old stage is 16, accounting for 9.71\%.
In this paper, we focus on how to leverage the high reliability of SLC and TLC to reduce the number of read retries in QLC flash memory and thus improve overall QLC SSD read performance at the expense of a small amount of capacity.

\section{RARO Design}
In this section, we first present an overview of the general design of RARO and a new principle for conversion and migration. Then, we give the flash mode switching strategy and dynamic data migration mechanism based on data hotness and cell reliability. Finally, the whole framework workflow is introduced. 

\subsection{\textbf{Overview}}\label{AA}
\begin{figure*}[htbp]
    \centering
    \includegraphics[width=0.8\linewidth]{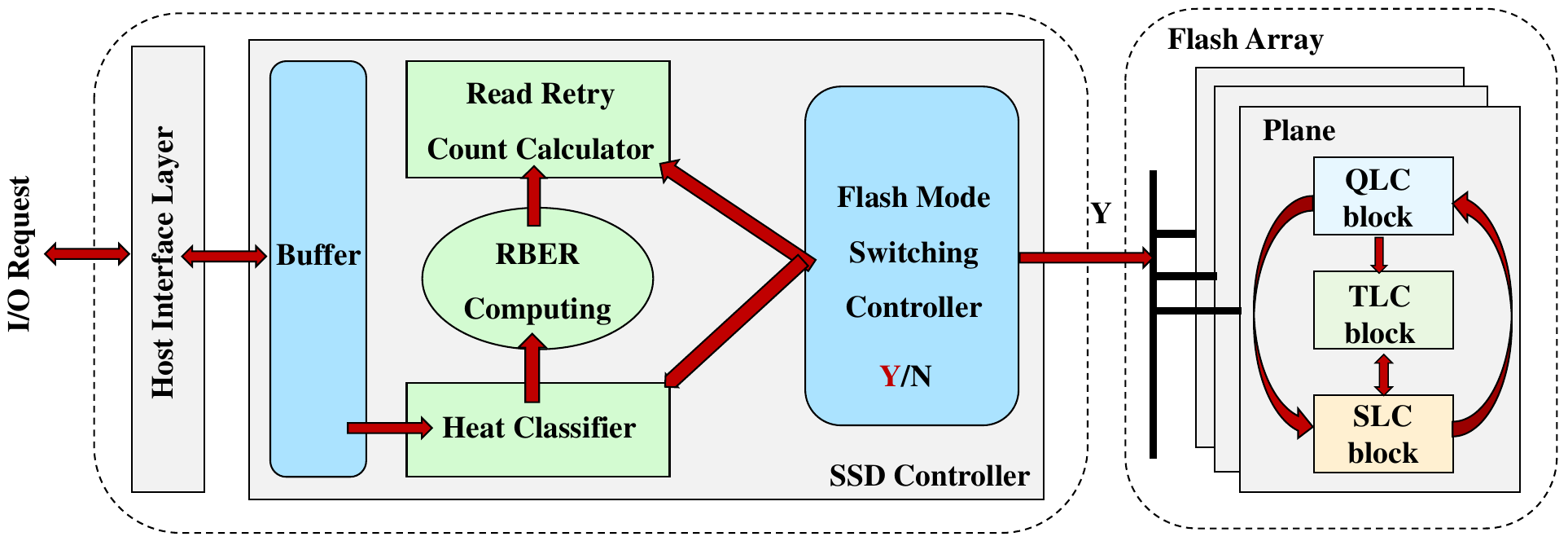}
    \caption{Architecture overview of RARO-SSD.}
    \label{fig7}
\end{figure*}

As shown in \Cref{fig7}, the RARO Method adds the following four main modules to the Flash Translation Layer (FTL) of the SSD: heat classifier, RBER computing, read retry count calculator, and flash mode translation controller.

The read retry count calculator is designed to calculate the retries for each flash page based on the RBER computing, which works in coordination with the heat classifier to place similar data in the same region. The mode conversion controller coordinates the dynamic reconfiguration between various NAND flash cell types, such as SLC, TLC, and QLC. 

\subsection{\textbf{Conversion and Migration Principles}}
The conventional hot/cold hierarchical storage strategy based on a single data access frequency has remarkable limitations in hybrid flash architectures.
In the early period of QLC NAND flash device usage, premature migration of high-access data to SLC/TLC modes will be costly in terms of wear and tear due to redundant erase operations, as the P/E cycles are low and the electronic escape has not yet accumulated.

With the surge of P/E cycles, the charge escape and read interference effects intensify, and the hot data stored in the reliability-degraded QLC flash memory cells will face the problem of severe read performance degradation due to the aggressive increase of read retry frequency. 
Hence, as shown in TABLE \ref{tab:migration}, this paper proposes a principle of dynamic data migration based on data temperature and read retry count awareness for a hybrid flash storage architecture.

\begin{table}[htbp]
\caption{Migration Principles}
\begin{center}
\begin{tabular}{|c|c|c|c|}
\hline
\textbf{NAND} & \textbf{\textit{Accesss Frequency} }& \textbf{\textit{Retry Count}} & \textbf{\textit{Mode Conversion}} \\
\hline
QLC & Hot & $\geq R_1 $ & QLC--SLC \\
\hline
QLC & Warm & $\geq R_2 (R_1 \leq R_2$) & QLC--TLC \\
\hline
TLC & Hot & $ \geq R_1$ & TLC--SLC \\
\hline
\end{tabular}
\end{center}
\label{tab:migration}
\end{table}

\subsection{\textbf{Dynamic Mode Switching Mechanism}}
According to the motivation section, the significant diffusion of read performance of flash memory cells of various densities due to different voltage distribution states can remarkably increase the overall read performance of SSD devices by placing the hot data with frequent read retries onto the lower-density flash memory cells with higher reliability.
Inspired by this, this paper describes a dynamic mode switching strategy (i.e., data is placed on flash memory in different modes by category) to coordinate the space allocation and state transition process of different types of flash memory in hybrid SSDs by combining the hotness of data and the number of read retries.

It is investigated that due to the limitation of the application scenarios, the industry does not have an interface involving the conversion from MLC to QLC flash memory for the time being, and hence, this paper considers the combination of the three types of flash memory, namely, SLC, TLC, and QLC.

\begin{figure}[htbp]
    \centering
    \includegraphics[width=6cm]{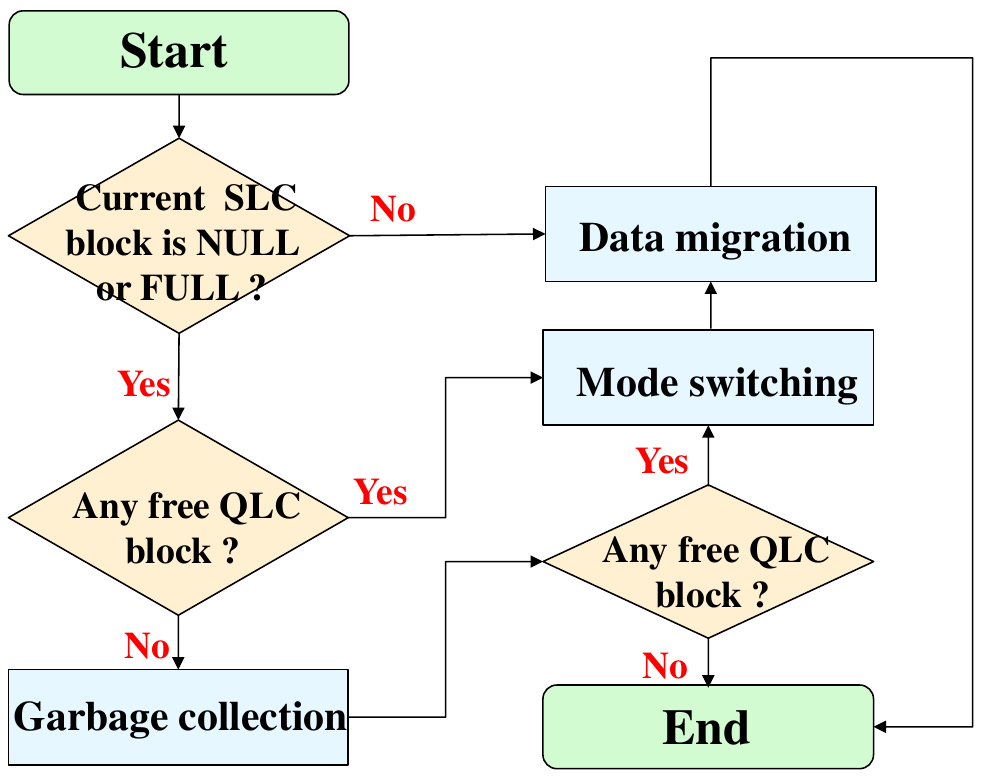}
    \caption{QLC to SLC Mode Switching workflow.}
    \label{fig}
\end{figure}

The cooperative optimization strategy between data access frequency and storage medium reliability is first established in a hybrid flash architecture as follows: 
\begin{itemize}
\item For 'hot data' with high-frequency access characteristics and the repeat number of reads exceeding the dynamic threshold $R_1$, a cross-level migration from QLC to SLC mode is performed to inhibit the reliability degradation caused by charge leakage; 
\item For 'warm data' with medium access intensity and the repeat number of reads exceeding $ R2 (R2 \geq R1) $ is migrated to TLC mode to balance latency and storage density;
\item For low-frequency access or read retries not reaching the threshold, continue to maintain QLC storage to relegate relocation expenditure.
\end{itemize}

\begin{figure}[htbp]
    \centering
    \includegraphics[width=8cm]{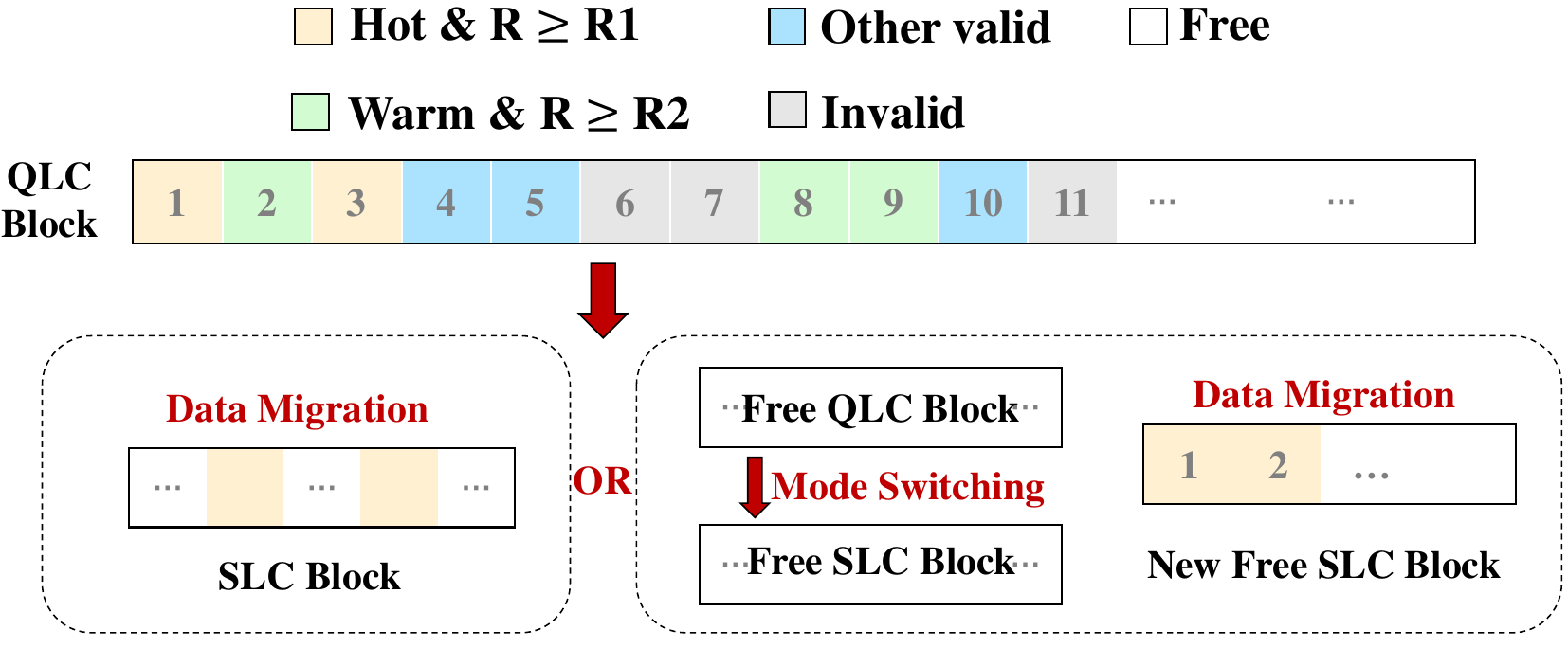}
    \includegraphics[width=8cm]{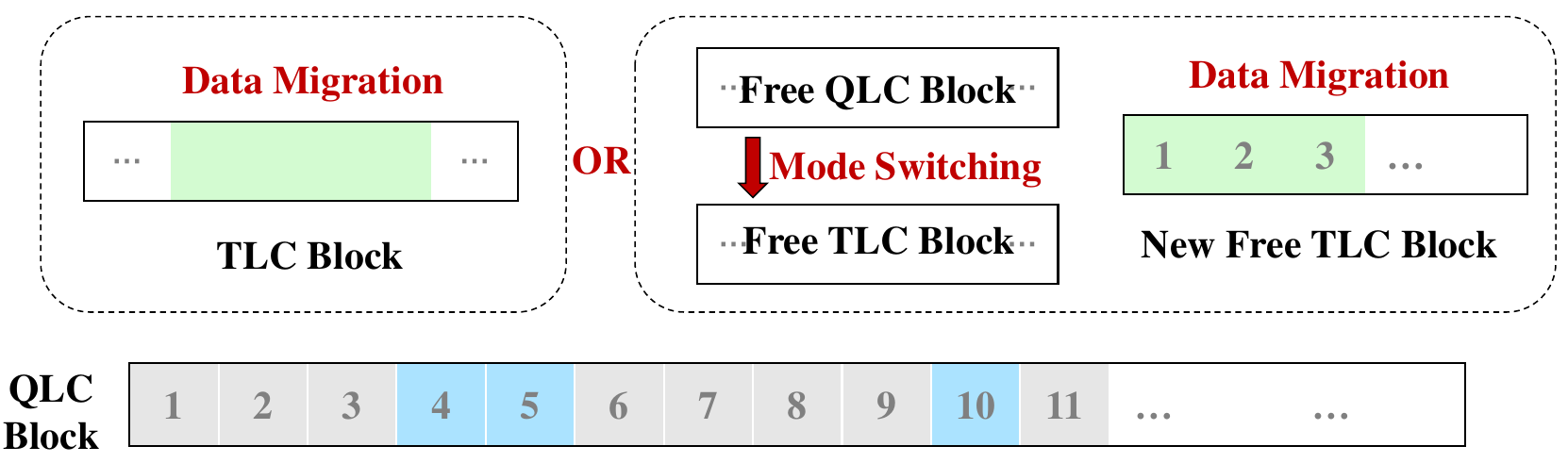}
    \caption{An example of QLC to SLC/TLC mode switching.}
    \label{fig}
\end{figure}

\begin{figure}[htbp]
    \centering
    \includegraphics[width=8cm]{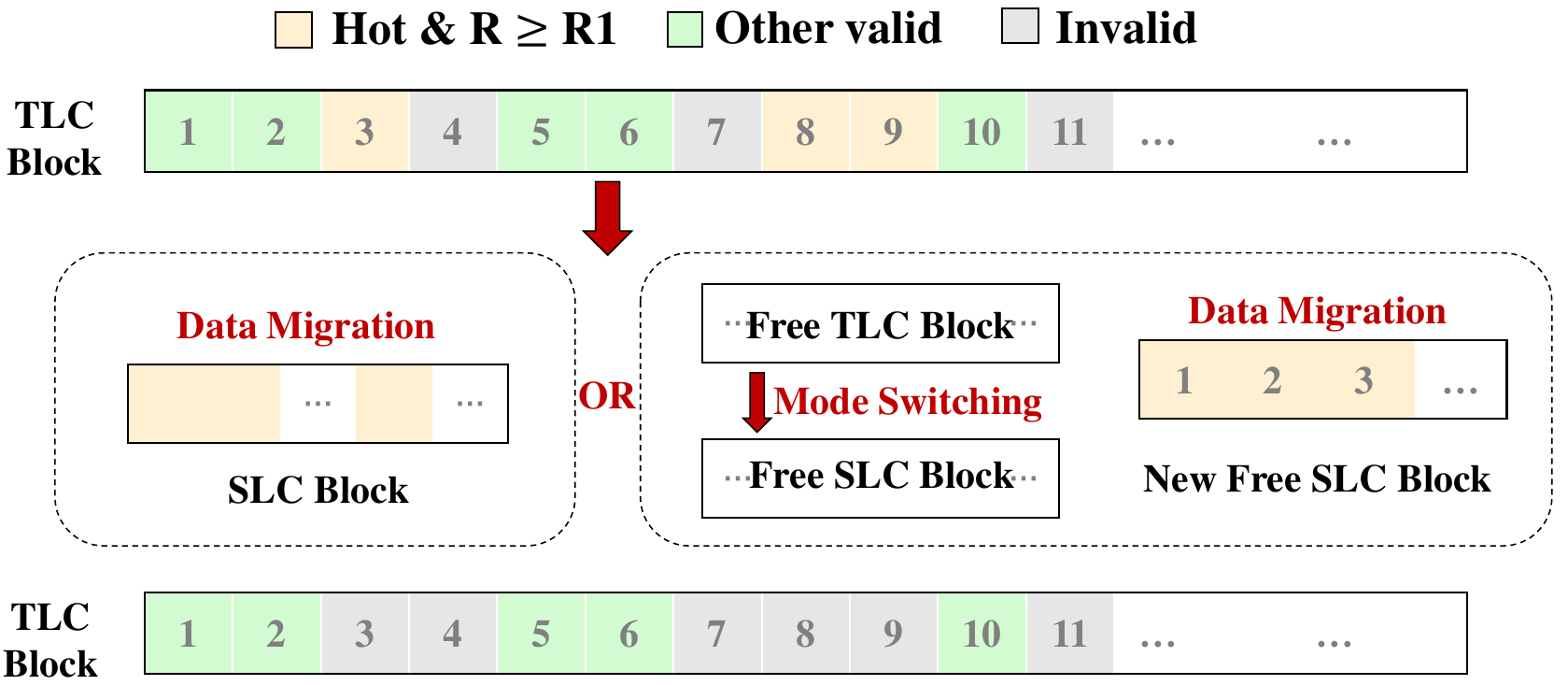}
    \caption{An example of TLC to SLC mode switching.}
    \label{fig}
\end{figure}

The configuration of the threshold parameter in this proposed study is essentially a multi-objective decision issue that requires a dynamic balancing mechanism between read performance gain and migration loss constraints. 
When the upper bound of the retry count threshold is set too high, the QLC storage tier will suffer from the cumulative effect of persistent read disturbances, triggering system-level performance degradation, which is contrary to the architectural design idea of improving read performance by reducing the number of read retries through schema conversion.
On the contrary, when the lower threshold parameter is set too low, the effectiveness of the retry count on hot data filtering will be weakened, while the non-essential migration operations will considerably increase the block erase and write fleets, which will accelerate the wear and tear of the storage units.
Specific parameter configurations will be quantitatively analysed in the experimental evaluation section with typical load scenarios.

\subsection{\textbf{Workflow Framework}}
Based on the dynamic data migration mechanism and capacity recovery policy, the whole working flow would be as follows.

This method constructs a three-stage data processing pipeline based on real-time analysis. Firstly, the hotness classifier determines the cold, warm, and hot attributes of the access frequency of the target physical page corresponding to the read request.
Secondly, it activates the read retry count calculator to compute the raw bit error rate (RBER) of the current page based on the aging model of the storage medium (background part), and then derives the theoretical read retry frequency.
Finally, the mode conversion controller decides whether to trigger the dynamic data migration by the metrics obtained in the first two steps.

To maintain the physical consistency of the storage architecture, the migration operation follows the principle of flash type alignment, i.e., taking the block as the smallest management unit to guarantee that all pages within the block remain uniform.
Through the full closed-loop processing flow of feature extraction-reliability assessment-decision execution, the dynamic matching of storage resources and access demand is achieved.

\begin{figure}[htbp]
    \centering
    \includegraphics[width=8cm]{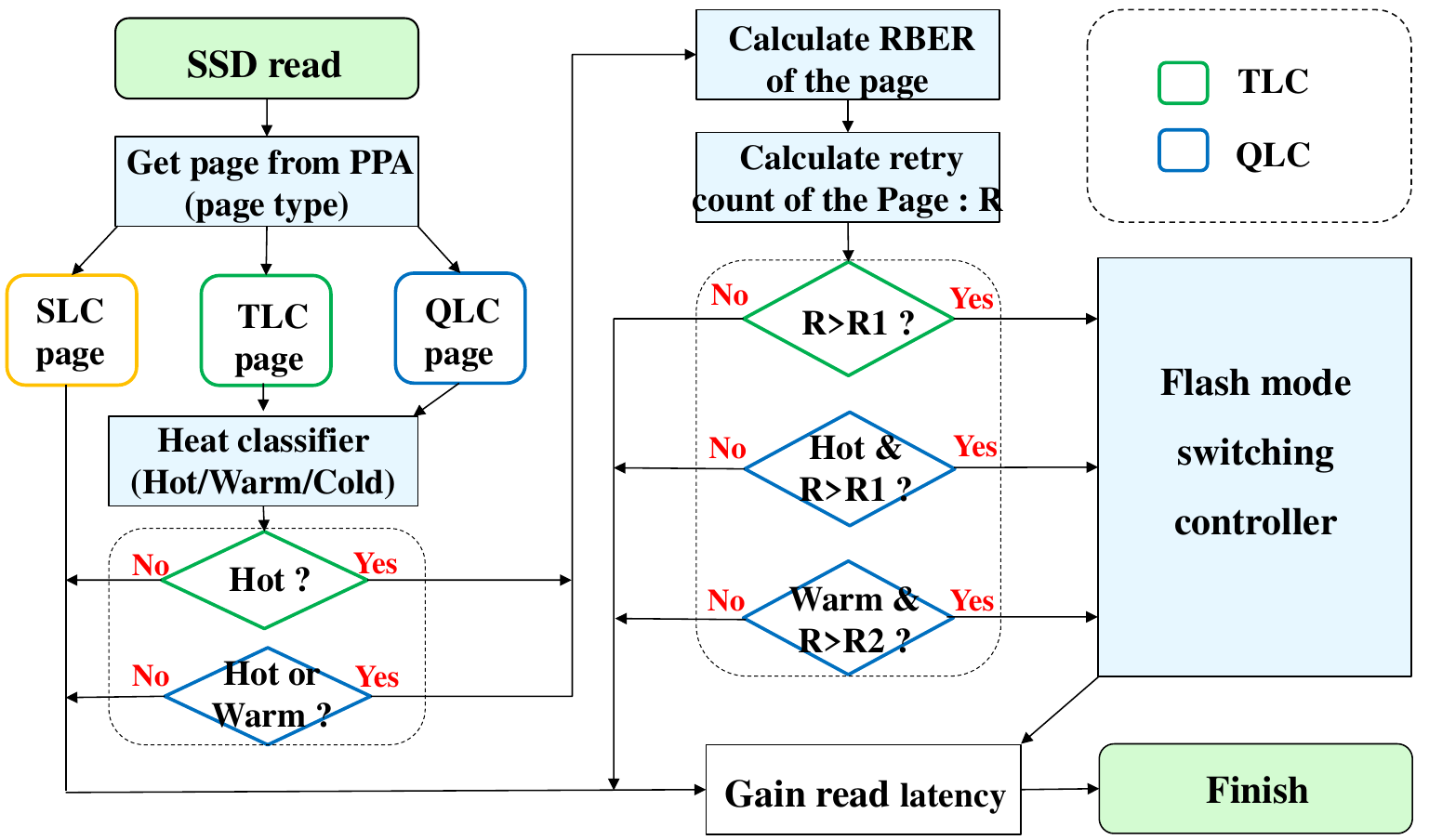}
    \caption{Workflow framework of RARO.}
    \label{fig}
\end{figure}

\subsection{\textbf{Further Discussion}}
The RARO approach proposed in this paper involves three mode transitions from high-density to low-density flash modes to reduce the read performance degradation problem due to a large number of read retries faced by QLC.
The data temperature involved in the mode conversion process for hybrid storage architectures varies dynamically, and the original hot data gradually loses its storage tiering justification as the access frequency decreases. 
Therefore, it is possible to similarly add the conversion from low-density to high-density flash storage mode based on the dynamically changing characteristics of the data heat lifecycle, bringing about an appropriate expansion of capacity.

\begin{figure}[htbp]
    \centering
    \includegraphics[width=8cm]{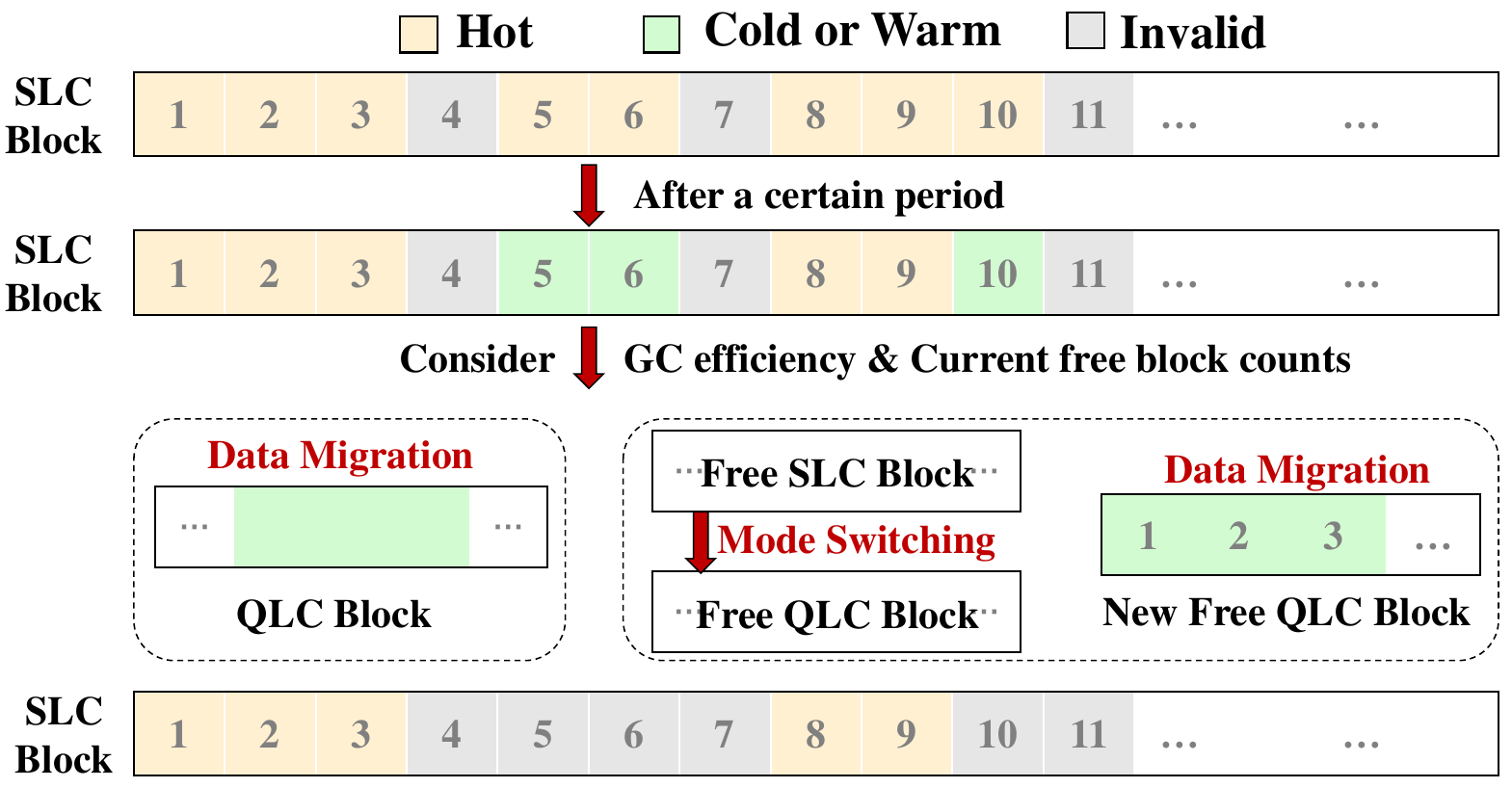}
    \caption{Temperature-based reclaim.}
    \label{fig12}
\end{figure}

For example, \Cref{fig12} shows that hot data previously migrated to SLC may become cold data after a while, and prolonged residence in high-performance storage media will impede the tiering migration path of newly generated hot data.
 If only periodic forced degradation and mode conversion to QLC are taken, not only will it produce significant write amplification, but it will also be difficult to cope with the risk of sudden changes in potential access patterns.
 Therefore, in this case, it is necessary to take into account the remaining space of the device, the efficiency of rubbish collection, and the user's writing demand.

\section{Experimental Result}
\subsection{\textbf{Experimental Setup}}\label{AA}
The experiments are conducted in FEMU\cite{li2018case}, a QEMU-based NVMe SSD emulator\cite{210518}. 
FEMU runs on a Linux server with a 40-core Intel(R) Xeon(R) CPU.
Table \ref{tab:configuration} lists the internal parallelism of the emulated hybrid SSDs, which are set according to the configuration of modern SSD products.
The read/program/erase latencies are shown in Table \ref{tab:latency}. 
Initially, the block types of the hybrid SSD are set to the QLC mode, with SLC and TLC blocks appearing when mode conversions are performed.

\begin{table}[htbp]
\caption{Configuration of Emulated SSD}
\begin{center}
\begin{tabular}{|c|c|c|c|}
\hline
\textbf{Configure} & \textbf{\textit{Value} }& \textbf{\textit{Configure}} & \textbf{\textit{Value}} \\
\hline
\# of channel & 2 & Pages per SLC block & 256 \\
\hline
LUNs per channel & 2 & Pages per TLC block & 768 \\
\hline
Planes per lun & 1 & Pages per QLC block & 1024 \\
\hline
Blocks per plane & 256 & Pages Size (KiB) & 16 \\
\hline
\end{tabular}
\end{center}
\label{tab:configuration}
\end{table}

\begin{table}[htbp]
\caption{Characteristics of SLC, TLC and QLC flash memories \cite{wei2023reinforcement,yoo2020reinforcement}}
\begin{center}
\begin{tabular}{|c|c|c|c|}
\hline
\textbf{NAND} & \textbf{SLC} & \textbf{TLC} & \textbf{QLC} \\
\hline
\textbf{\textit{Bits/cell}} & 1 & 3 & 4 \\
\hline
\textbf{\textit{Read latency}} & $\sim 20us $ & $\sim 66us $ & $\sim 140us $ \\
\hline 
\textbf{\textit{Write latency}} & $\sim 160us $ & $\sim 730us $ & $\sim 3102us $ \\
\hline
\textbf{\textit{Erase latency}} & $\sim 2ms $ & $\sim 3ms$ & $\sim 10ms$\\
\hline
\textbf{\textit{P/E cycles}} & 100000 & 3000 & 1000\\
\hline
\end{tabular}
\end{center}
\label{tab:latency}
\end{table}

To evaluate the optimisation of the RARO method for QLC SSD read performance, we implemented RARO in FEMU SSDs and compared it with the Base scheme and Hotness.
The base scheme represents a multiple read retry approach for QLC SSDs without considering the characteristics of different flash modes, and Hotness indicates the conversion and migration between the three flash memory modes SLC-TLC-QLC based only on temperature. And Hotness was chosen for comparison to verify the validity of the conversion and migration principle proposed in this paper.

SSDs use page-level address mapping algorithms and greedy rubbish collection algorithms that are widely used in modern SSDs. In our experiments, most of the workloads were generated using the Flexible I/O tester (FIO) \cite{Fio}, where the logical addresses of the I/O requests conform to different Zipf distributions. All the workloads have a data set as large as 8 GB.

\subsection{\textbf{Overall Read Performance Comparison}}
The read retry thresholds in this section are selected based on the conclusions of the sensitivity tests in the third part of the experiment. 
Since there is no heat variation for sequential reads, we only compare the random read performance of the three methods under workloads with different access skews.

\begin{figure}[htbp]
    \centering
    \includegraphics[width=8cm]{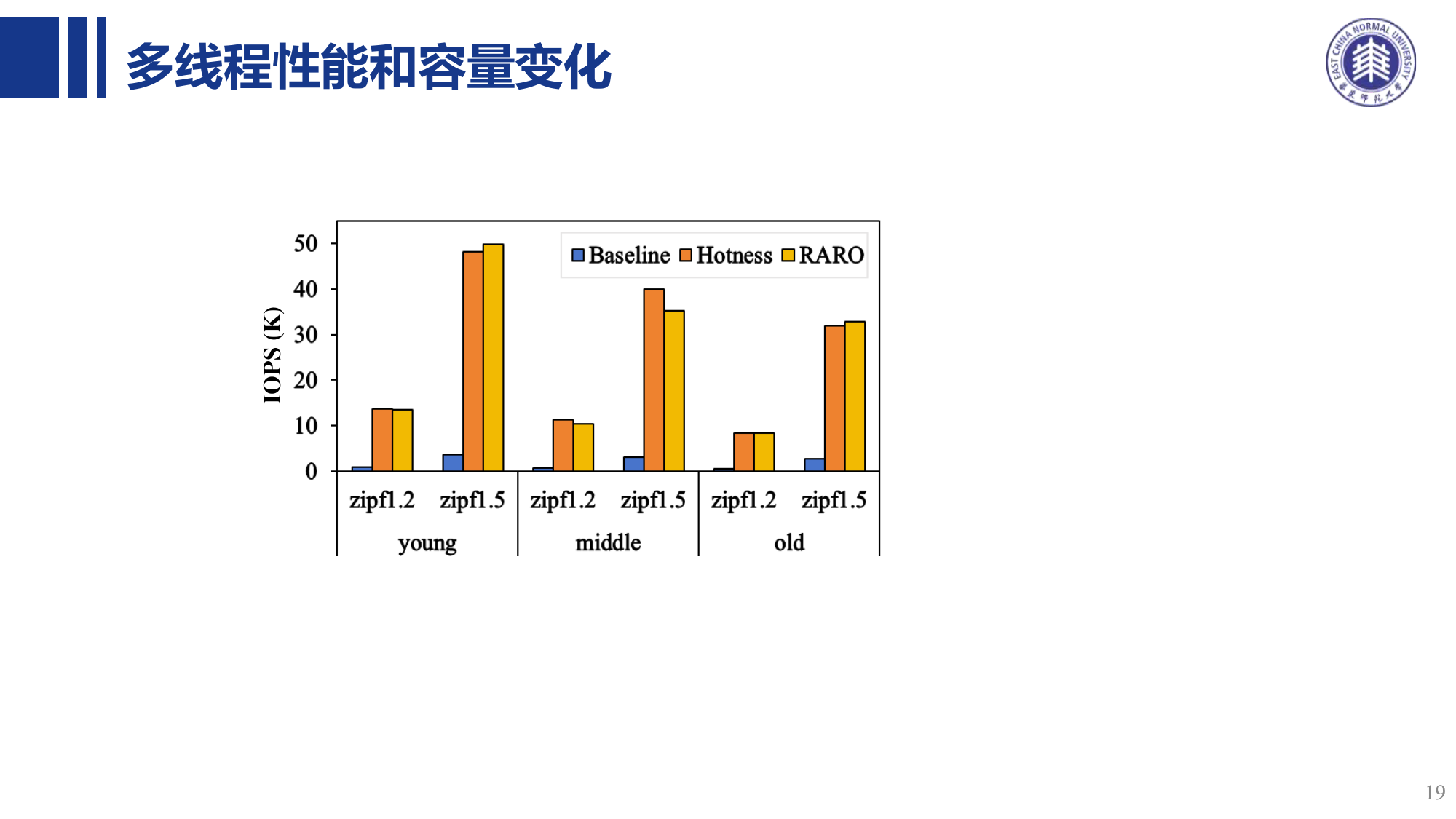}
    \caption{Comparison of random read IOPS for three methods under different workloads with four threads.}
    \label{fig13}
\end{figure}

\begin{figure}[htbp]
    \centering
    \includegraphics[width=8cm]{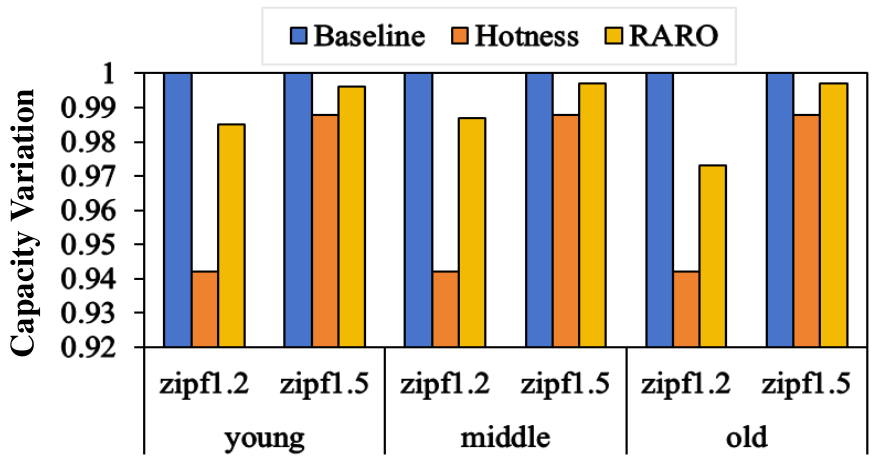}
    \caption{Comparison of capacity change for three methods under different workloads with four threads.}
    \label{fig14}
\end{figure}

As can be seen from \Cref{fig13}, our proposed RARO method has the highest random read performance of 13.5, 12.8, and 13.2 times of the baseline scheme for all three stages of flash memory: young, middle and old of flash memory, respectively. This is due to the fact that a large amount of data located in high-latency flash mode is migrated to low-latency flash mode, which reduces the number of read retries located in high-density flash mode and improves the overall read performance of the device.

\Cref{fig14} shows the changes in final and initial capacity for the three stages. It can be seen that compared to the Hotness scheme, our proposed RARO scheme optimises the capacity loss by 74.1\%, 77.6\% and 53.4\% for young, middle and old ages under zipf 1.2 workload respectively, and 66.7\%, 75\% and 75\% for young, middle and old ages, respectively, under zipf 1.5 workload. The RARO approach proposed in this paper has great advantages in capacity loss relying on the proposed conversion principle, not only based on temperature but also need to reach the retry count threshold.

\begin{figure}[htbp]
    \centering
    \includegraphics[width=8cm]{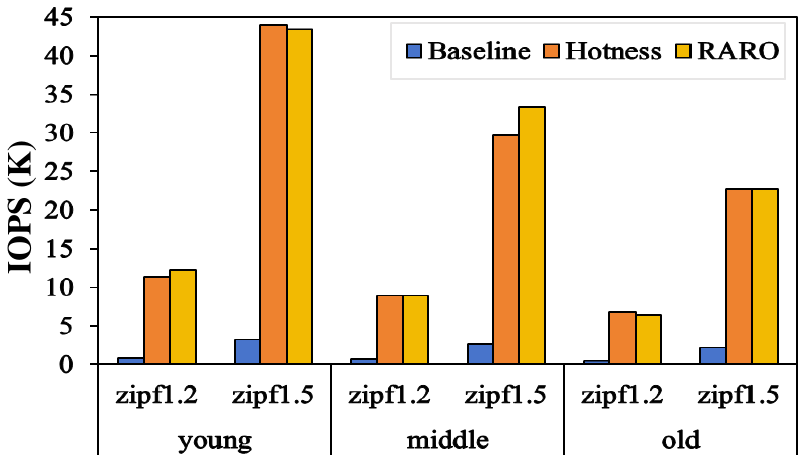}
    \caption{Comparison of random read IOPS for three methods under different workloads with single threads.}
    \label{fig15}
\end{figure}

\begin{figure}[htbp]
    \centering
    \includegraphics[width=8cm]{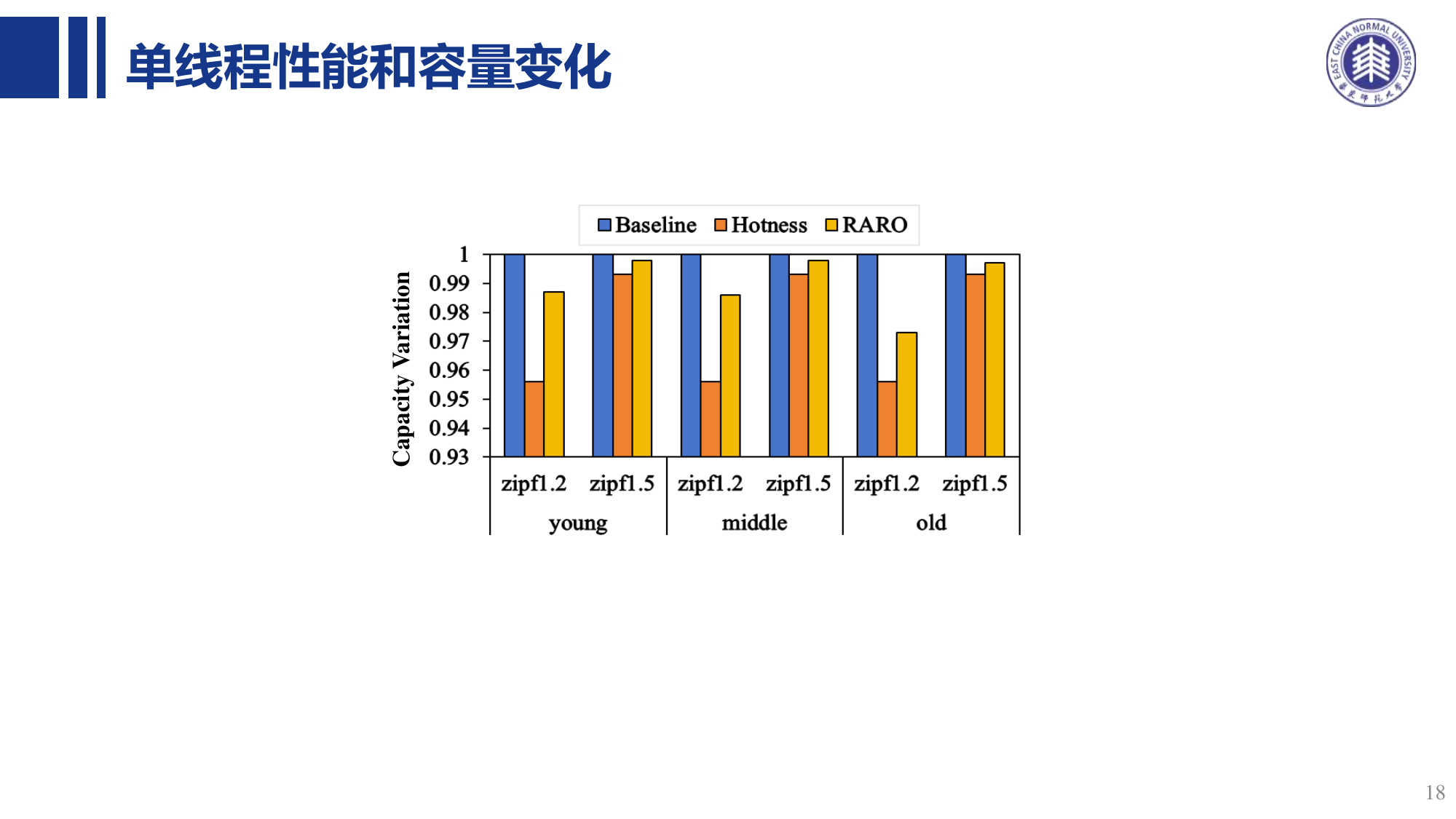}
    \caption{Comparison of capacity change for three methods under different workloads with single threads.}
    \label{fig16}
\end{figure}

\Cref{fig15} and \Cref{fig16} analyse the single-threaded case, and draw essentially similar conclusions to the multi-threaded one. In the single-thread random read scenario, our proposed RARO scheme optimises the capacity loss of young, middle and old by 70.4\%, 68.2\% and 38.6\%, respectively, under zipf 1.2, and 71.4\%, 71.4\% and 57.1\%, respectively, under zipf 1.5, as compared to Hotness's scheme. 

Taken together, our scheme significantly improves the random read performance of QLC SSDs and is able to reduce a significant amount of data migration overhead and capacity loss while maintaining essentially the same read performance as the Hotness scheme.

\subsection{\textbf{Sensitivity Analysis}}
There are two read retry count threshold parameters in RARO: $R_1$ and $R_2$. To understand the characteristics of these two parameters, we conducted a sensitivity study to show their impact on the overall read performance and capacity of the device.
According to the read retries distribution in section \uppercase\expandafter{\romannumeral3}, the number of page read retries of the QLC flash memory in the experiment is centrally located between 4 and 9 in the young stage, between 7 and 12 in the middle stage, and between 11 and 16 in the old age stage, so the read retries threshold $R_2$ varies according to the ranges of these three stages.
The TLC flash memory in the experiment is converted from QLC flash memory, and it is tested under typical workloads and found that the number of read retries does not exceed 1, so the read retry threshold $R_1$ is selected as 1.

\begin{figure}[htbp]
    \centering
    \includegraphics[width=8cm]{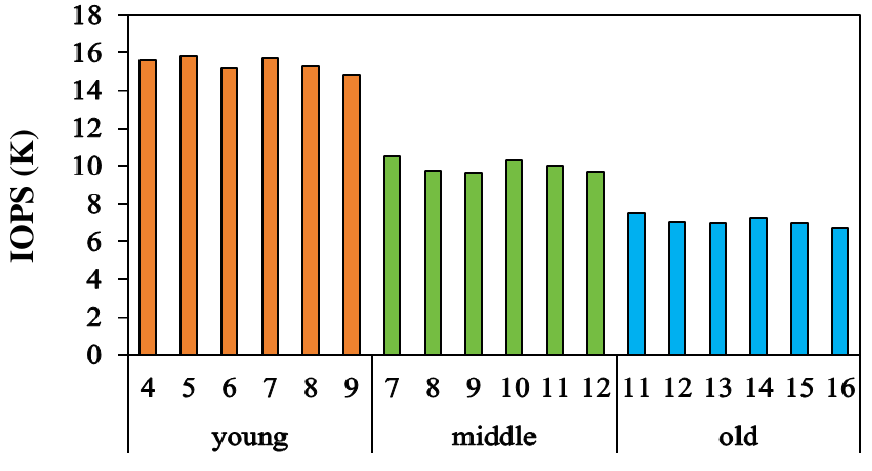}
    \caption{Comparison of the impact of read retry thresholds R2 on read performance under young/middle/old phases.}
    \label{fig}
\end{figure}

\begin{figure}[htbp]
    \centering
    \includegraphics[width=8cm]{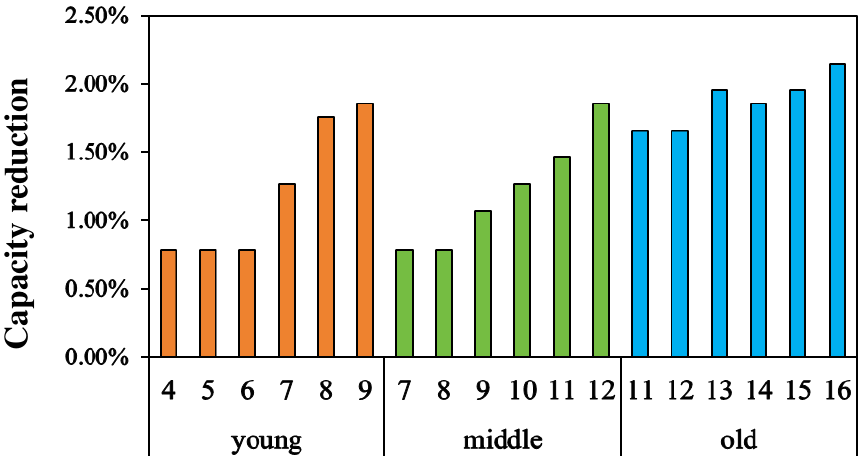}
    \caption{Comparison of the capacity reduction by read retry threshold R2 under young/middle/old phases.}
    \label{fig}
\end{figure}

It can be seen from the figure that roughly the larger the threshold $R_2$ is set, the more the overall capacity is lost. This may be because when the threshold $R_2$ increases, the conversion from QLC to SLC increases while the conversion from QLC to TLC decreases.
Taking the reading performance and capacity loss into account, the threshold $R_2$ was set to 5, 7, and 11 in the young, middle, and old stages, respectively. In practice, the thresholds $R_1$ and $R_2$ may need to be adjusted according to workload characteristics.

\section{Conclusion}
In this paper, we propose a read optimisation method, RARO, based on SLC-TLC-QLC three-mode conversion to improve the read performance of QLC SSDs at the expense of a small amount of device capacity. 
We model the data with diverse characteristics based on the number of read retries, and place the data in different flash modes with various reliability for reading, while coordinating SSD internal status.
Experimental results show that our RARO scheme can effectively improve system bandwidth without incurring significant capacity loss and with low overhead.



 


\bibliographystyle{unsrt}
\bibliography{ref}

\begin{thebibliography}{10}

\bibitem{luo-hyghdensity}
Longfei Luo, Dingcui Yu, Hang Li, Yunpeng Song, Yina Lv, Edwin H.-M. Sha, and Liang Shi.
\newblock Revisiting trim on high-density flash-based hybrid storage systems.
\newblock {\em IEEE Transactions on Computer-Aided Design of Integrated Circuits and Systems}, 43(5):1618--1622, 2024.

\bibitem{ssd665p}
{Intel ssd 665p series}, 2019.

\bibitem{crucialP1}
{Micron crucial p1 product}, 2018.

\bibitem{luo2023performance}
Longfei Luo, Shicheng Li, Yina Lv, and Liang Shi.
\newblock Performance and reliability optimization for high-density flash-based hybrid ssds.
\newblock {\em Journal of Systems Architecture}, 136:102830, 2023.

\bibitem{8942140}
Bingzhe Li, Chunhua Deng, Jinfeng Yang, David Lilja, Bo~Yuan, and David Du.
\newblock Haml-ssd: A hardware accelerated hotness-aware machine learning based ssd management.
\newblock In {\em 2019 IEEE/ACM International Conference on Computer-Aided Design (ICCAD)}, pages 1--8, 2019.

\bibitem{wei2023reinforcement}
Qian Wei, Yi~Li, Zhiping Jia, Mengying Zhao, Zhaoyan Shen, and Bingzhe Li.
\newblock Reinforcement learning-assisted management for convertible ssds.
\newblock In {\em 2023 60th ACM/IEEE Design Automation Conference (DAC)}, pages 1--6. IEEE, 2023.

\bibitem{10693661}
Yumiao Zhao, Jiancong Zheng, Gangyong Jia, Liang Shi, and Congming Gao.
\newblock Hook: A pattern locality guided prefetch with enhanced read performance for hybrid ssds.
\newblock In {\em 2024 13th Non-Volatile Memory Systems and Applications Symposium (NVMSA)}, pages 1--6, 2024.

\bibitem{li2018case}
Huaicheng Li, Mingzhe Hao, Michael~Hao Tong, Swaminathan Sundararaman, Matias Bj{\o}rling, and Haryadi~S Gunawi.
\newblock The $\{$CASE$\}$ of $\{$FEMU$\}$: Cheap, accurate, scalable and extensible flash emulator.
\newblock In {\em 16th USENIX Conference on File and Storage Technologies (FAST 18)}, pages 83--90, 2018.

\bibitem{Fio}
{Fio: flexible io tester}.

\bibitem{7056062}
Yu~Cai, Yixin Luo, Erich~F. Haratsch, Ken Mai, and Onur Mutlu.
\newblock Data retention in mlc nand flash memory: Characterization, optimization, and recovery.
\newblock In {\em 2015 IEEE 21st International Symposium on High Performance Computer Architecture (HPCA)}, pages 551--563, 2015.

\bibitem{10.1145/3224432}
Yixin Luo, Saugata Ghose, Yu~Cai, Erich~F. Haratsch, and Onur Mutlu.
\newblock Improving 3d nand flash memory lifetime by tolerating early retention loss and process variation.
\newblock 2(3), 2018.

\bibitem{6657034}
Yu~Cai, Onur Mutlu, Erich~F. Haratsch, and Ken Mai.
\newblock Program interference in mlc nand flash memory: Characterization, modeling, and mitigation.
\newblock In {\em 2013 IEEE 31st International Conference on Computer Design (ICCD)}, pages 123--130, 2013.

\bibitem{8327033}
Yixin Luo, Saugata Ghose, Yu~Cai, Erich~F. Haratsch, and Onur Mutlu.
\newblock Heatwatch: Improving 3d nand flash memory device reliability by exploiting self-recovery and temperature awareness.
\newblock In {\em 2018 IEEE International Symposium on High Performance Computer Architecture (HPCA)}, pages 504--517, 2018.

\bibitem{kim2019design}
Bryan~S Kim, Jongmoo Choi, and Sang~Lyul Min.
\newblock Design tradeoffs for $\{$SSD$\}$ reliability.
\newblock In {\em 17th USENIX Conference on File and Storage Technologies (FAST 19)}, pages 281--294, 2019.

\bibitem{lv2023mgc}
Yina Lv, Liang Shi, Qiao Li, Congming Gao, Yunpeng Song, Longfei Luo, and Youtao Zhang.
\newblock Mgc: Multiple-gray-code for 3d nand flash based high-density ssds.
\newblock In {\em 2023 IEEE International Symposium on High-Performance Computer Architecture (HPCA)}, pages 122--136. IEEE, 2023.

\bibitem{cho2024aero}
Sungjun Cho, Beomjun Kim, Hyunuk Cho, Gyeongseob Seo, Onur Mutlu, Myungsuk Kim, and Jisung Park.
\newblock Aero: Adaptive erase operation for improving lifetime and performance of modern nand flash-based ssds.
\newblock In {\em Proceedings of the 29th ACM International Conference on Architectural Support for Programming Languages and Operating Systems, Volume 3}, pages 101--118, 2024.

\bibitem{park2021reducing}
Jisung Park, Myungsuk Kim, Myoungjun Chun, Lois Orosa, Jihong Kim, and Onur Mutlu.
\newblock Reducing solid-state drive read latency by optimizing read-retry.
\newblock In {\em Proceedings of the 26th ACM International Conference on Architectural Support for Programming Languages and Operating Systems}, pages 702--716, 2021.

\bibitem{210518}
Huaicheng Li, Mingzhe Hao, Michael~Hao Tong, Swaminathan Sundararaman, Matias Bj{\o}rling, and Haryadi~S. Gunawi.
\newblock The {CASE} of {FEMU}: Cheap, accurate, scalable and extensible flash emulator.
\newblock In {\em 16th USENIX Conference on File and Storage Technologies (FAST 18)}, pages 83--90, Oakland, CA, February 2018. USENIX Association.

\bibitem{yoo2020reinforcement}
Sangjin Yoo and Dongkun Shin.
\newblock Reinforcement $\{$Learning-Based$\}$$\{$SLC$\}$ cache technique for enhancing $\{$SSD$\}$ write performance.
\newblock In {\em 12th USENIX Workshop on Hot Topics in Storage and File Systems (HotStorage 20)}, 2020.

\end{thebibliography}

\end{document}